\documentclass{elsart3p}

\usepackage{graphicx}

\usepackage{amssymb}

\newcommand{\beq}{\begin{equation}}
\newcommand{\eeq}{\end{equation}}
\newcommand{\beqa}{\begin{eqnarray}}
\newcommand{\eeqa}{\end{eqnarray}}

\begin{document}

\begin{frontmatter}



\title{Charge inhomogeneity coexisting with large Fermi surfaces}


\author[a]{A. Di Ciolo}
\author[a]{M. Grilli}
\author[a]{J. Lorenzana}
\author[b]{G. Seibold}

\address[a]{SMC-INFM,ISC-CNR and Dipartimento di Fisica, Universit\`a di 
Roma ``La Sapienza'', Piazzale Aldo Moro 2, I-00185 Roma, Italy}
\address[b]{Institut f\"ur Physik, BTU Cottbus, PBox 101344,
         03013 Cottbus, Germany}

\begin{abstract}
We discuss how stripes in cuprates can be compatible with a Fermi-liquid-like
Fermi surface and, at the same time, they give rise to a one-dimensional-like pseudo Fermi surface in the momentum distribution function. 
\end{abstract}

\end{frontmatter}


There is substantial evidence that charge and spin inhomogeneities are 
present in underdoped superconducting cuprates in the form 
of stripe or checkerboard textures\cite{review}. 
Close to a quantum critical point the detection of order may strongly depend on the
typical timescales of experimental probes\cite{andergassen}.
 In particular two different Fermi surfaces (FS) were identified in 
Ref.~\cite{zho99Zho01} in an underdoped 
${\rm La_{2-x}Sr_xCuO_4}$ (LSCO) sample by integrating the spectral function
in an energy window: 
$
n({\Omega,\bf k})\equiv \int_{{-\Omega}}^{{\Omega}} d \omega A({\bf k},\omega)f(\omega)
$.

Using  a small energy window $\Omega=$15meV, $n(\Omega,{\bf k})$ 
roughly corresponds to  $A({\bf k},E_F)$, and Zhou and collaborators
obtained the usual large Fermi surface of a uniform Fermi liquid.

Using a large energy window, instead, ($\Omega=100 ~\sim 250$meV)
Zhou and collaborators obtained an approximate map of the  momentum
distribution function  $n({\bf k})= n(\infty ,\bf k)$ and defined the
Fermi surface by the locus of the largest gradients in $n({\bf k})$. 
This second procedure
identified a FS made of crossed straight lines as one would 
expect in the presence of well-formed magnetically ordered stripes. 
 Indeed 
this Fermi surface has been computed in LDA+U\cite{ani04} and 
is in good agreement with the results of  Ref.~\cite{zho99Zho01} including
fine details.  

For a Fermi liquid, treating data with great accuracy at $T=0$, 
the two approaches should give the
same FS, but in practice, if the energy distribution curves
are dominated by strong incoherent high-energy excitations, the quasiparticle
contribution to $n({\bf k})$ may be hard to detect and a pseudo FS
may be determined from the specific momentum structure of excitations at high
binding energy.  A similar finding would be obtained by observing the FS of a heavy
fermion with an energy resolution larger or smaller than the coherence energy
scale (of the order of the Kondo energy scale $\sim T_K$). If the resolution is
very good the small jump $\sim T_K/E_F$ in $n({\bf k})$ identifies a large FS, while
with poor resolution this small jump is missed and a pseudo FS with
small volume would be
observed corresponding to the FS of the decoupled conduction
band\cite{kur00}. 

We propose that a similar scenario is realized in cuprates. 
Low-energy quasiparticles close to the Fermi energy are slow 
coherent excitations which 
form on long timescales and therefore see a time-averaged environment of spin
and (possibly) of charge. In this case the environment seen by the quasiparticles 
is nearly homogeneous and a large (non-interacting) FS is recovered from their
dispersions. As in heavy fermions the contribution of quasiparticles
to $n({\bf k})$ can be very modest. Instead 
  $n({\bf k})$  will be dominated by 
the incoherent spectral weight which is
determined by short time effects, i.e. by a seemingly magnetically 
ordered environment. 

Based on this  physical picture 
we propose a simple procedure to compute the pseudo FS and
the true FS. In the first case one performs a 
mean-field computation of stripes as in
Refs.~\cite{ani04,lor02bsei04a}. These solutions have long-range order
in the charge and the spin sector and basically corresponds to a
snapshot of the fluctuating stripes. The obtained spectral weight 
should be identified with the incoherent spectral weight of the system
and therefore give an estimate of $n({\bf k})$ in which coherent
effects have been neglected. This explains the good agreement found.

At low energy however this charge and spin texture will fluctuate 
allowing for the formation of low energy quasiparticles. 
 In order to estimate the resulting FS one simply performs 
a mean-field computation averaging over the degrees of freedom that
fluctuate. If both charge and spin fluctuate, one trivially obtains the
usual non-interacting Fermi surface.  It is more likely that at  
doping $x\sim 1/8$ commensurate charge order is present but magnetic order is not
present since
charge order breaks a discrete symmetry whereas spin order breaks a
continuous symmetry and is much more sensitive to quantum
fluctuations\cite{note}. 
How this may be consistent with the observed large free-electron-like FS? 
Obviously a strong charge modulation would strongly affect the FS shape, but 
small charge modulation (as experimentally observed\cite{abb05})
could still be compatible with ARPES experiments and could produce
specific observable signatures. To test this idea we have computed the
FS constraining the charge modulation to the one obtained in the
magnetically ordered stripes (which is very small) while  constraining 
the spins to be in a paramagnetic state. 

Our computations are performed in the Hubbard model, where the strong correlation
appropriate for the cuprates is treated by the
Gutzwiller approximation.
The constraint in the charge is implemented
through a Lagrange multiplier. It is interesting that without the
constraint paramagnetic stripes are not found. This can be traced back
to the poor treatment of the  magnetic energy contribution
in the Gutzwiller approximation at large $U$ and for a paramagnetic state.

 The result is reported in Fig. 1 for a 
harmonic bond-centered (BC) charge modulation. The spectral intensity
has been symmetrized for the $x-y$ exchange by summing the solution
for stripes along $x$ and along $y$.
\begin{figure}[tbp]
$$\includegraphics[width=5 cm]{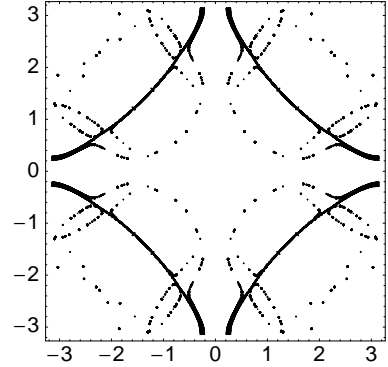}$$
\caption{Spectral weight $A({\bf k},E_F)$ in the Brillouin zone 
 for a symmetrized
$(4\times 1 + 1 \times 4)$ charge modulation in the Hubbard
model with  $U/t=9$, $t'=-0.2$, and  hole doping $x=\frac{1}{8}$. 
The charge is modulated with a difference between maxima and minima
$\Delta n=0.1$.}
\label{FS} 
\end{figure}

Owing to the small charge modulation,  the paramagnetic CO solutions 
still produce a large FS close to that expected from band-structure
calculations. Nevertheless distinctive features of the CO are also 
observable, which arise from the crossing of tenuous shadow bands in the
regions near the M points of the FS. Interestigly we found similar
features in the experimental FS's of Ref. \cite{zho99Zho01}.
We therefore conclude that a weak charge modulation is fully compatible
with the experimental observation of large FS's for the slow 
coherent quasiparticle excitations.


\begin{thebibliography}{1}
\bibitem{note}
Magnetic long range order has been detected in Nd-LSCO for which nodal
  quasiparticles have been observed, the ordered moment however corresponds to
  a minute fraction of the total moment which may be indicative of large
  patches of disordered regions.
\end{thebibliography}

\end{document}